\documentclass[a4paper,11pt]{article}
\usepackage{pos}

\usepackage{xspace}
\usepackage{mathrsfs}
\usepackage{amsfonts}
\usepackage{array}
\usepackage{epsfig}
\usepackage{rotating}
\usepackage{multirow}

\usepackage{amsmath}    
\usepackage{verbatim}   
\usepackage{color}      
\usepackage{colordvi}
\usepackage{bm}

\usepackage{subfigure}  
\usepackage{url}

\usepackage{pstricks,rotating, graphicx}
\usepackage{cancel}

\usepackage{tablefootnote}  

\renewcommand{\logo}{\relax}

\title{
Top-quark pole mass extraction at NNLO accuracy
}
\ShortTitle{
Top-quark pole mass extraction at NNLO accuracy
}

\author[a]{S.~Alekhin}
\author*[b,a]{M.V.~Garzelli}
\author[c]{J.~Mazzitelli}
\author[a]{S.-O.~Moch}
\author[a]{O.~Zenaiev}

\affiliation[a]{II. Institut f\"ur Theoretische Physik, Universit\"at Hamburg \\
	Luruper Chaussee 149, D--22761 Hamburg, Germany}
\affiliation[b]{Department of Theoretical Physics, CERN,  
CH--1211 Geneva 23, Switzerland}
\affiliation[c] {Paul Scherrer Institut, CH--5352 Villigen, Switzerland}

\emailAdd{sergey.alekhin@desy.de}
\emailAdd{maria.vittoria.garzelli@desy.de}
\emailAdd{javier.mazzitelli@psi.ch}
\emailAdd{sven-olaf.moch@desy.de}
\emailAdd{oleksandr.zenaiev@desy.de}

\abstract{We describe our recent NNLO QCD extraction of the top-quark pole mass  from fits to experimental data on total inclusive and normalized (multi)-differential cross sections for $t\bar{t} + X$ hadroproduction, using as input various modern PDF~+~$\alpha_s(M_Z)$ sets. 
We find top-quark mass values compatible among each other and with the PDG 2024 preferred value.}

\note[]{Report number: CERN-TH-2024-225}

\FullConference{12th Large Hadron Collider Physics Conference  (LHCP2024)\\
3-7 June 2024 \\
Boston, USA \\}

\begin{document}
\maketitle

\section{Introduction}
The top-quark mass $m_t$, as well as other quark masses, are parameters in the Standard Model. Comparing experimental cross sections for $t\bar{t} +X$ hadroproduction, as measured at the TeVatron and the Large Hadron Collider (LHC), to theory predictions in a well-defined mass renormalization scheme, allows for the extraction of the top-quark mass value. We follow such a path in Ref.~\cite{Garzelli:2023rvx}, working at next-to-next-to-leading order (NNLO) QCD accuracy and focusing on the top-quark pole mass. Here we summarize the methodology and the most important results from this study. 

\section{Methodology}
\begin{figure}
\begin{center}
\includegraphics[width=0.49\textwidth]{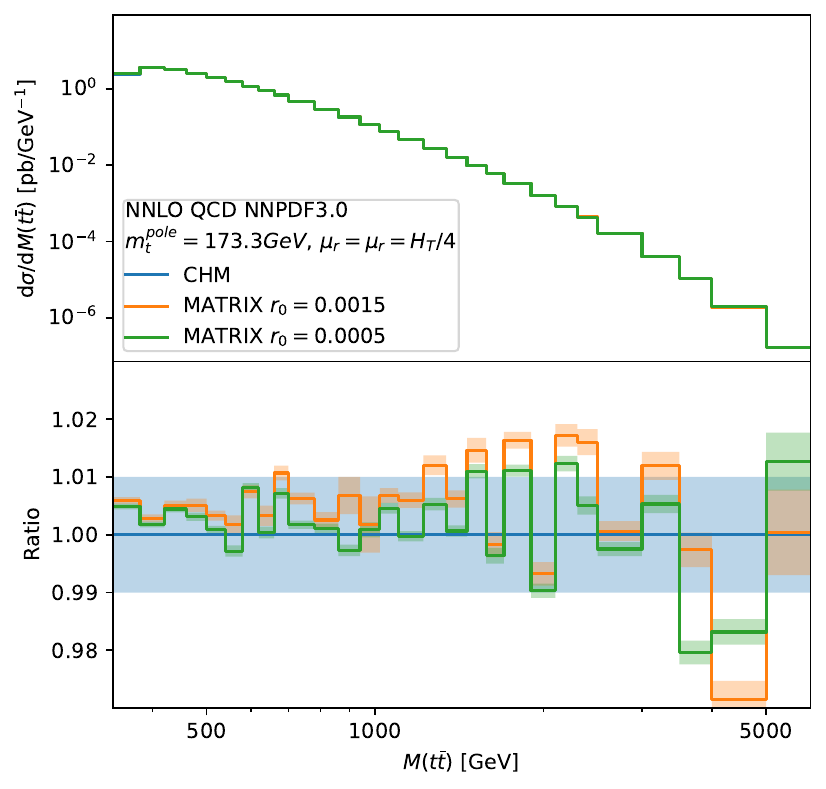}
\includegraphics[width=0.49\textwidth]{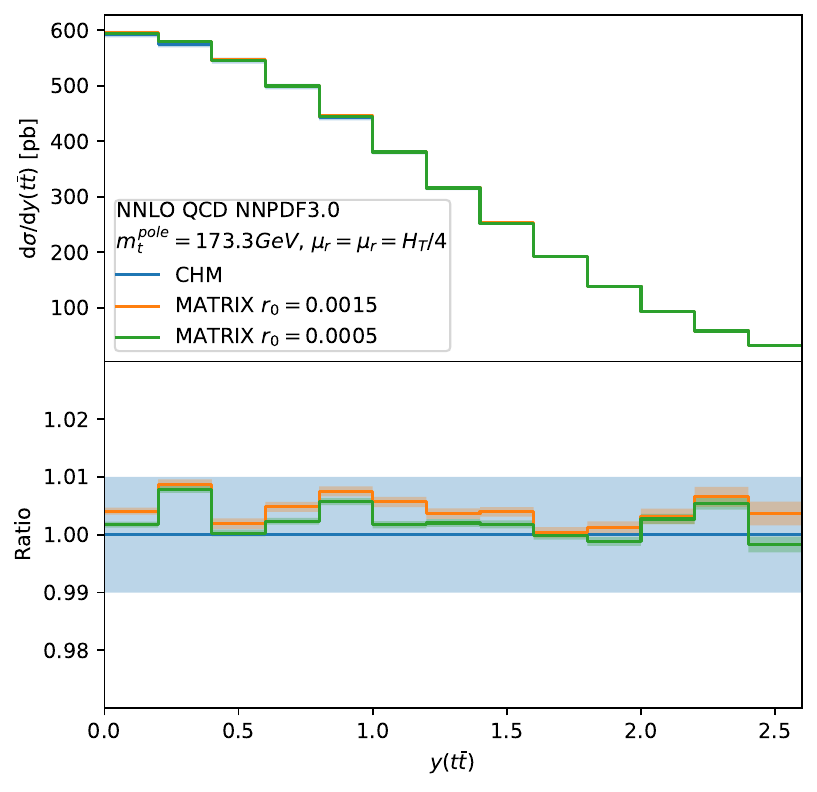}
\caption{\label{fig:power} Comparison of predictions for the invariant mass (left panel) and the rapidity (right panel) distributions of top-quark pairs in $pp \rightarrow t\bar{t} + X$ at $\sqrt{S}$~=~13~TeV obtained by our version of \texttt{MATRIX} with $r_0$~=~0.15\% (orange) and with $r_0$~=~0.05\% (green), with the STRIPPER predictions of Ref.~\cite{Czakon:2016dgf} (blue), here labelled as CHM. The bands refer to numerical uncertainties in the various computations.  $r_0$ is a cut-off on $q_T/M(t\bar{t})$, used in $q_T$ subtraction. The exact cross sections correspond to the limit $r_0 \rightarrow 0$ ($q_T \rightarrow 0$). 
}
\end{center}
\end{figure}

As a first ingredient for the fit, we compute theory predictions for cross sections at 
NNLO accuracy in QCD
by means of \texttt{MATRIX}~\cite{Grazzini:2017mhc}+~\texttt{PineAPPL}~\cite{Carrazza:2020gss}.
We started, in particular, from the \texttt{MATRIX} version that was first prepared for producing the predictions in Ref.~\cite{Catani:2019hip}, on which we operated a number of technical modifications that allow for a better handling of the memory and of jobs in parallelized intensive computations, as required to reach a statistical accuracy well below 1\% in bins of differential distributions. We interface this \texttt{MATRIX} version to \texttt{PineAPPL}, which allow to create interpolation grids for handling easily and fast the convolution with different PDF~+~$\alpha_s(M_Z)$ sets, as well as renormalization and factorization scale variation, as soon as a first set of predictions has been produced. On the other hand, different \texttt{MATRIX} runs are required for different top-quark mass input values. We work in the on-shell top-quark mass renormalization scheme. \texttt{MATRIX} is based on the $q_T$-subtraction formalism, which, in turns, is closely related to $q_T$-resummation~\cite{Catani:2014qha, Catani:2018mei}. This formalism
has been succesfully applied to the computation of total and (multi-)differential cross sections for a number of processes up to NNLO QCD. $q_T$-subtraction is a non-local infrared (IR) singularity subtraction method, implying that a certain class of power corrections should be included to restore exact NNLO cross sections. A comparison of our predictions to those obtained by the local IR subtraction method STRIPPER~\cite{Czakon:2014oma, Czakon:2011ve, Czakon:2010td}, shows that these power corrections, that we are neglecting, are not large, at least for the distributions included in this study, see Fig.~\ref{fig:power}.

A second main ingredient of the fit are suitable experimental data. 
A plethora of data  on $t\bar{t} + X$ production exist, especially at the LHC. As for total cross sections, we choose to use all the datasets appearing in the LHCTop Working Group summary plot~\cite{lhctopwg}, whereas, in case of differential distributions, we choose to use the datasets that satisfy all the following constraints: 1) normalized cross sections are provided, to decrease the effects of the lack of information on correlations of uncertainties between different analyses within an experimental group or from different experimental groups; 2) results are unfolded to the parton level: we expect that the experimentalists already take care of all systematics in their unfolding from particle to parton level; 3) the dataset information on systematical uncertainties and bin-by-bin correlations of the latter is publicly available; 4) the results are provided as  distributions
in the invariant mass $M(t\bar{t})$ of the $t\bar{t}$ pair, or double-differential distributions in $M(t\bar{t})$ and rapidity $y(t\bar{t})$, considering that $M(t\bar{t})$ is particularly sensitive to the top-quark mass. We observe also that, since we use normalized cross sections, instead of absolute ones, this sensitivity shows up in a non-negligible way in all bins, even at large $M(t\bar{t})$, and not only in the threshold region. 

In our theory predictions, we do not include neither the effects of the resummation of logarithms associated to the emission of soft real gluons nor the effect of the exchange of Coulomb gluons between the $t$ and $\bar{t}$ quarks, which might become large close to threshold (see, e.g., Ref.~\cite{Garzelli:2024uhe} and references therein). We expect that these effects are not particularly relevant in our study, considering that the present width of the experimental bins is quite large, implying that threshold effects may play only a minor role in the fit considering bins with those sizes. Of course, as soon as experimental results with thinner binning close to threshold will become available, it will become important to include
resummation effects in the corresponding theory predictions. 

The top-quark mass value is extracted through a least-square fit, comparing data to theory predictions computed as described above using as input four different modern NNLO PDF~+~$\alpha_s(M_Z)$ sets: ABMP16~\cite{Alekhin:2017kpj}, CT18~\cite{Hou:2019efy}, MSHT20~\cite{Bailey:2020ooq} and NNPDF4.0~\cite{NNPDF:2021njg}. 
In the covariance matrix used to build the $\chi^2$ we include experimental statistical, systematical correlated and uncorrelated data uncertainties, as well as theory uncertainties. The latter include PDF uncertainties and 
a numerical uncorrelated uncertainty of 1\% in each bin, related to the lack of power corrections in our {\texttt{MATRIX}} computations, as well as to Monte Carlo numerical integration considering a limited number of phase-space points and to the use of {\texttt{PineAPPL}} interpolation grids in place of exact calculations. On the other hand, renormalization and factorization scale uncertainties are not included in the $\chi^2$, but we repeat the fit for different ($\mu_r$, $\mu_f$) combinations in a seven-point scale variation procedure. The envelope of the results gives rise to a separate uncertainty on the top-quark mass.

We compute theory predictions and the $\chi^2$ values for several fixed values of $m_t^{\rm{pole}}$. Close to the minimum, the $\chi^2$ turns out to show a reasonable parabolic shape as a function of $m_t^{\rm{pole}}$ for all PDF~+~$\alpha_s(M_Z)$ sets considered. In each case, we fit the three lowest values of $\chi^2$ as a function of $m_t^{\rm{pole}}$ through a parabola, and we quote as best-fit value the minimum of the parabola. 

\section{Results of the fit and discussion}

\begin{figure}
\begin{center}
 \includegraphics[width=0.49\textwidth]{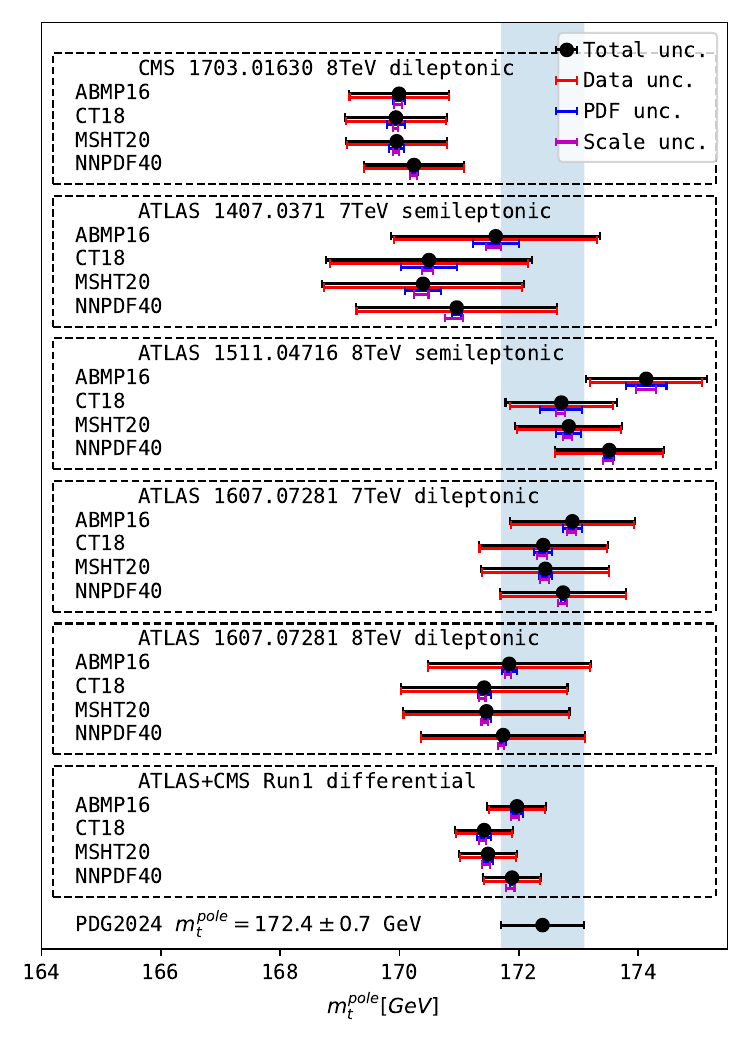}
 \includegraphics[width=0.49\textwidth]{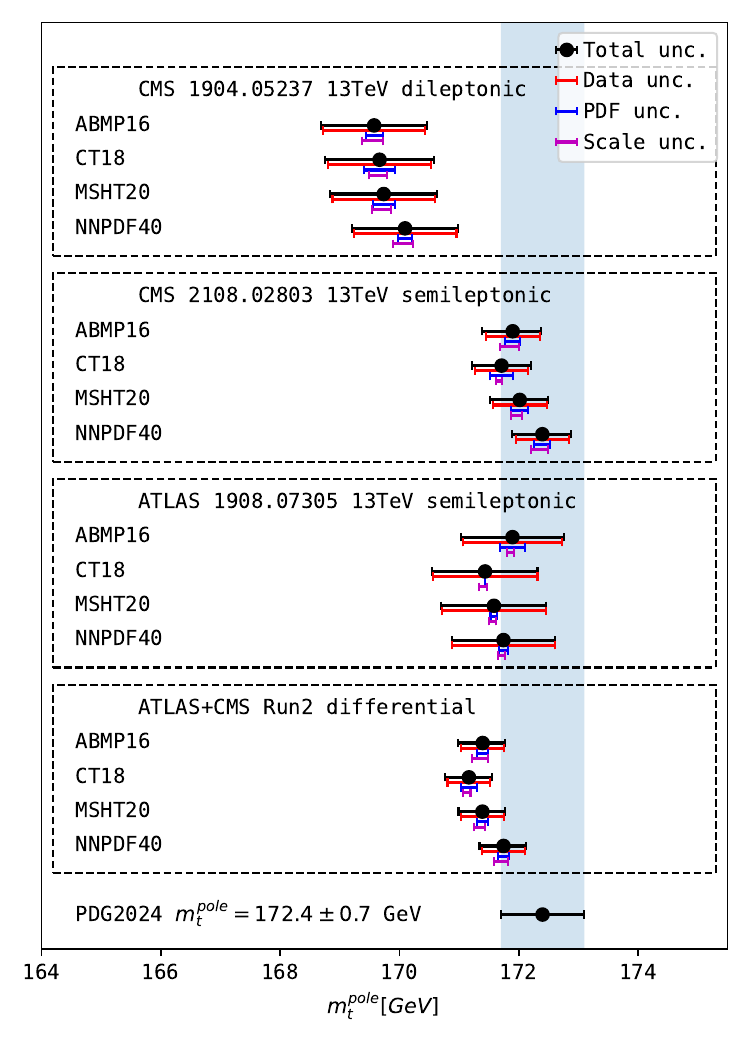}
 \includegraphics[width=0.49\textwidth]{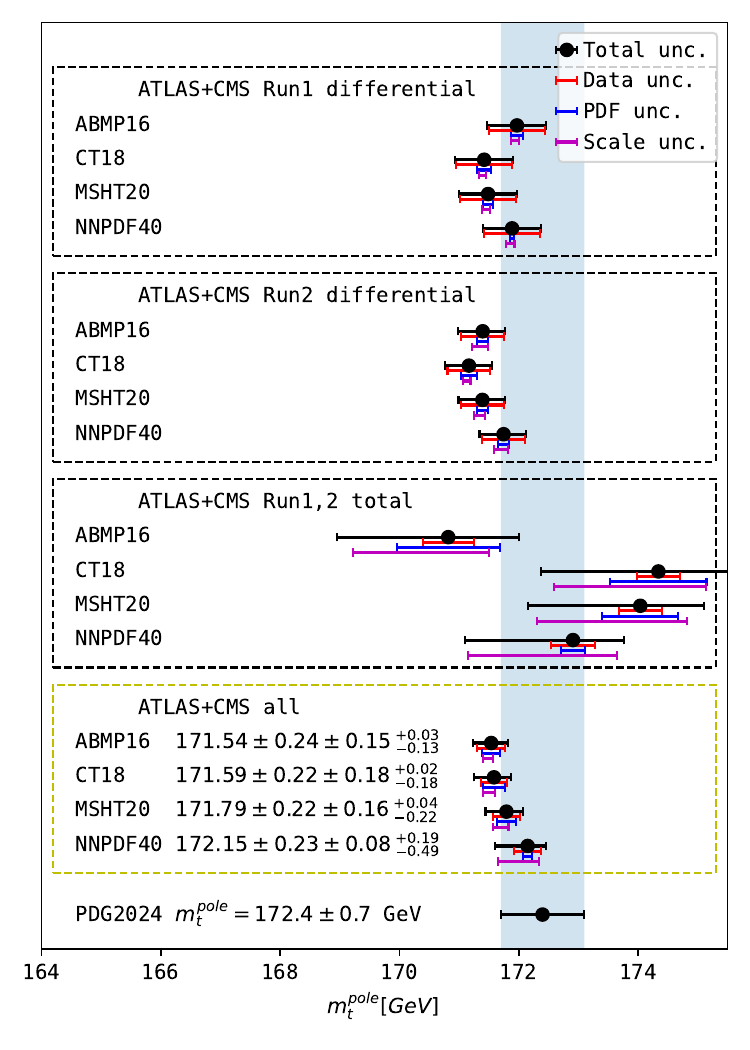}\\
\caption{\label{fig:plotsfit} $m_t^{\rm{pole}}$ values extracted using datasets of normalized (multi-)differential cross sections obtained in Run I (left upper panel) and Run II (right upper panel), as well $m_t^{\rm{pole}}$ values from a global analysis of Run 1 + Run 2 differential and total inclusive cross sections (lower panel). The light-blue band in all panels corresponds to the PDG 2024 $m_t^{\rm{pole}}$ value. In the inset in the lowest panel, the best-fit $m_t^{\rm{pole}}$ values from the global fit, together with the data, PDF and scale uncertainties, are reported for each of the considered PDF~+~$\alpha_s(M_Z)$ sets adopted as input.
}
\end{center}
\end{figure}

For each PDF~+~$\alpha_s(M_Z)$ set, we report the results of the fits to each group of datasets separately, distinguishing Run 1 differential datasets, Run 2 differential datasets, the ensemble of total cross-section datasets, as well as the ``global'' fit, 
including all datasets.
By comparing the results in the panels of Fig.~\ref{fig:plotsfit}, it is evident that 1) Run 2 differential datasets have much larger constraining power that Run 1 datasets, and they dominate in the global fit; 2) total cross-section datasets are not particularly relevant, and the results for top-quark mass best-fit value obtained by using them as standalone strongly depend on the $\alpha_s(M_Z)$ value, considering the high degree of correlation  existing among the $m_t$ value, PDFs and $\alpha_s(M_Z)$ in total cross sections; 3) on the other hand, in normalized cross sections, the effects of $\alpha_s(M_Z)$ greatly cancel in the ratio of numerator and denominator. It then turns out, in part also as a consequence of this fact, that the best-fit top-quark mass values extracted using as input a same normalized dataset and different PDF~+~$\alpha_s(M_Z)$ sets, are compatible among each other, which helps making our top-quark mass fit robust; 4) our results turn out to be compatible with the PDG 2024 $m_t^{\rm{pole}}$ value~\cite{ParticleDataGroup:2024cfk}.

Our quoted uncertainties include those from the data, as well as theory uncertainties due to PDF and scale variation. At present the experimental and theory uncertainty components have roughly a similar size.  While a part of the analyses with the full Run-2 integrated luminosity are still under completion, we are looking forward to updated high-statistics experimental data 
using Run 3 and High-Luminosity LHC, which will help to further reduce the statistical uncertainties in our fit. Further theory efforts will then be necessary to match the reduced experimental uncertainty.

We have released publicly our \texttt{MATRIX + PineAPPL} grids of predictions of $t\bar{t}+X$ double-differential cross sections to facilitate further fits by other theory and experimental groups.

\acknowledgments

The work of S.A., M.V.G. and S.-O.M. has been supported in part by the Deutsche Forschungsgemeinschaft through the Research Unit FOR 2926, {\it Next Generation pQCD for Hadron Structure: Preparing for the EIC}, project number 40824754.
The work of O.Z. has been supported by the {\it Philipp Schwartz Initiative} of the Alexander von
Humboldt foundation


%

\bibliographystyle{JHEP} 
\bibliography{ttab_lhcp}

\end{document}